\documentclass[a4paper,11pt]{article}
\pdfoutput=1 

\usepackage{jcappub} 

\usepackage[T1]{fontenc} 

\usepackage[normalem]{ulem}
\usepackage{array,multirow,graphicx}
\usepackage{dcolumn}
\usepackage{txfonts}
\usepackage{newlfont}
\usepackage{times}
\usepackage{bm}
\usepackage{hyperref}
\usepackage[figtopcap]{subfigure}
\usepackage{float}
\usepackage{ulem}
\usepackage{xcolor}
\usepackage[T1]{fontenc}
\usepackage{ae,aecompl}
\usepackage{nccmath}
\usepackage{soul}

\newcommand{\LCDM}{\Lambda \textrm{CDM}}
\newcommand{\fTCDM}{f(T)\text{-}\textrm{CDM}}
\newcommand{\sgmI}{1\sigma}
\newcommand{\sgmII}{2\sigma}

\graphicspath{{./figs/}}

\title{Toward a Concordance Teleparallel Cosmology I: Background Dynamics}

\author{Mahmoud Hashim,}%
\author[1]{Waleed El Hanafy,\note{Corresponding author.}}
\author{Alexey Golovnev}%
\author{and Amr A. El-Zant}%

\affiliation{Centre for Theoretical Physics, The British University in Egypt, P.O. Box 43, El Sherouk City, Cairo 11837, Egypt}

\emailAdd{mahmoud@aims.ac.za}
\emailAdd{waleed.elhanafy@bue.edu.eg}
\emailAdd{alexey.golovnev@bue.edu.eg}
\emailAdd{amr.elzant@bue.edu.eg}

\abstract{%
Assuming a spatially flat universe, we study the cosmological viability of an infrared corrected teleparallel gravity model, which  accounts for late acceleration by weakening gravity at later times on cosmological distances. The theory does not introduce
any additional free parameters into the cosmological model, as is commonly the case
with modified gravity based cosmologies.
This feature renders the cosmological model statistically comparable, on equal
footing, with $\LCDM$. In this context,
using recent cosmological observations --- Pantheon supernova Type Ia, Hubble constant $H_0$, Baryon acoustic oscillation, redshift space distortions, Big Bang nucleosynthesis and the cosmic microwave background constraint on the decoupling acoustic scale ---  we show that, although the exponential infrared-corrected  gravity and $\LCDM$ are  physically different, they are phenomenologically and statistically equivalent. However, the former  is more adept at fitting accurately
determined observational  constraints while decreasing
the $H_0$ tension without worsening the $S_8$ tension. This calls for full examination of the empirical viability
of the theory at the linear perturbation level,
which is the subject of paper II.
}

\begin{document}
\maketitle
\section{Introduction}\label{Sec:1}
Cosmological observations of Supernovae Type Ia (SNIa) distances embodied the first direct evidence for a shift from cosmic deceleration to acceleration a few billion years ago \cite{Riess:1998AJ, Perlmutter:1999ApJ}. This phenomenon can be accounted for through the introduction of a dark energy (DE) component, with negative pressure, into Friedmann equation. The simplest description invokes a cosmological constant $\Lambda$, with an equation of state (EoS) fixing the ratio of pressure to density to
$w_{DE}=-1$, in the field equations of general relativity (GR). Adding this to a pressureless cold dark matter (CDM) component defines the dark sector of the $\LCDM$ model. Although the model suffers theoretical problems, such as fine tuning \cite{Weinberg:1988cp,Carroll:2000fy}, it precisely fits a variety of
cosmological and astrophysical observations.  The simplicity and empirical success of
$\LCDM$ has lead to its wide acceptance,  fundamental problems regarding the origin of the cosmological constant notwithstanding.

However, recent observations, with unprecedented accuracy, provide some evidence of possible physics beyond $\LCDM$. A major problem involves the current value of the Hubble parameter $H_0$. The cosmic microwave background (CMB) observations by Planck (base-$\LCDM$) infer $H_0=67.4 \pm 0.5$ km/s/Mpc \citep{Aghanim:2018eyx}, while its most recent value according to direct local measurements using distance ladder methods (expanded sample of Milky Way Cepheids + Gaia EDR3 parallaxes) is $H_0=73.2 \pm 1.3$ km/s/Mpc, as measured by Riess et. al. 2020 with 1.8\% precision \citep[hereafter R20]{Riess:2020fzl}.
This is in 4.2$\sigma$ tension with Planck (for other similar measurements see also \citep{Riess:2016jrr, Riess:2018jrr, Riess:2018byc,Riess:2019cxk}).


The $H_0$ value inferred by Planck, assuming $\LCDM$, is supported by an independent dataset that combines the baryon acoustic oscillation (BAO) and  big bang nucleosynthesis (BBN) with clustering and weak lensing from the Dark Energy Survey (DES) \citep{Abbott:2017smn}. On the other hand, the late universe measurement R20 is supported by non distance ladder methods \citep{Wong:2019kwg, Pesce:2020xfe}. And the same conclusion has been achieved by using anchors other than Cepheids to calibrate the SNIa distance ladder \citep{Huang_2020,Freedman_2019}. In general, the tension between the early and the late universe of the $H_0$ measurements is at the 4$\sigma$--to--6$\sigma$ level for different combinations of datasets (\citep{Verde:2019ivm}, see also \citep{Riess:2020sih}).

The amplitude of matter fluctuations, $\sigma_8$, provide  additional evidence of tension between the early and the late universe measurements. This is often referred to as $S_8\equiv \sigma_8 \sqrt{\Omega_m/0.3}$ tension. Here, the CMB measurements imply $S_8 = 0.834 \pm 0.016$ \citep{Aghanim:2018eyx}, as inferred by Planck base-$\LCDM$, while cosmic shear base-$\LCDM$ observations from late universe give $S_8=0.745 \pm 0.039$ (as measured by Kilo Degree Survey~\citep[hereafter KiDs-450]{Hildebrandt:2016iqg}); and $S_8 = 0.737^{+0.040}_{-0.036}$ (as measured by KiDs+VIKING-450 \citep[hereafter KV-450]{Hildebrandt:2018yau}). These are about 2--2.3$\sigma$ lower than Planck. The preference for $\sigma_8$ values lower than Planck are obtained using redshift space distortion (RSD) observations~\citep{Macaulay:2013swa,Barros:2018efl}.

One may interpret the $H_0$ tension as a discrepancy in two measurements.
One (early universe) measure involves the sound horizon at radiation drag $r_{drag}$.
This being the case,
one may resort to changes in early universe physics to ease the tension
by reducing the sound horizon \citep{Cuesta:2014asa,Lemos:2018smw,Aylor:2018drw,Arendse:2019hev}. However, it has been shown that  classical extensions of $\LCDM$ ---  such as allowing for more parameters
by varying the number and masses of neutrinos --- do not solve the $H_0$ tension on their own. They also worsen the $\sigma_8$ tension (\citep{Aghanim:2018eyx}; see also \citep{Guo:2018ans}). On the other hand, in the extended 12 parameter space, when a dark energy equation of state $w<-1$ and neutrino species other than the standard ones are allowed to vary simultaneously, there is no preference for increasing $N_{eff}$ when $H_0$ tension is addressed \citep{DiValentino:2016hlg,DiValentino:2017zyq,DiValentino:2019dzu}.
Self interacting neutrinos provide a better framework for solving both tensions simultaneously. It remains a challenge to construct and verify viable models with requirements beyond standard model physics with very large couplings \citep{Kreisch:2019yzn, Vagnozzi:2019ezj}.
Localized energy injection around the matter-radiation equality epoch
has also been invoked to reduce the sound horizon and increase the Hubble constant by introducing early dark energy \citep{Poulin:2018cxd, Agrawal:2019lmo}. However, these models suffer fine tuning problems at eV scales, and also lead to severe scale dependent changes in the matter spectrum, which worsen the $\sigma_8$ tension \citep{Hill:2020osr}. They furthermore shift some standard $\LCDM$ parameters \citep{Smith:2019ihp}, in particular the spectral index $n_s$ and the physical baryon density
$\Omega_b h^2$ ($h = H_0/100$ km/s/Mpc).

A closed universe scenario was also suggested to solve some internal inconsistencies between high-$\ell>800$ and low-$\ell<800$ observations in the Planck data \citep{Handley:2019tkm,DiValentino:2019qzk}. However, it has been shown that when BAO is combined with Planck all parameters are enforced to flat $\LCDM$ even by considering the extension to 12 parameter space, which suggests even fundamentally radical extensions to the standard model need to keep the basic phenomenological elements of its success \citep{DiValentino:2020hov}. At the same line, it has been shown that the combination of Planck data with full-shape galaxy power spectrum \citep{Vagnozzi:2020zrh} and cosmic chronometers \citep{Vagnozzi:2020dfn} can be used to break the geometrical degeneracy, whereas the apparent preference for a closed universe from Planck disappears.

Proposals to reduce the $H_0$ tension through
modifications of late universe physics include those invoking
interacting dark matter \citep{DiValentino:2017iww},
emergent dark energy \citep{Li:2019ypi,Pan:2019hac} and modified gravity \citep{Nunes:2018xbm,El-Zant:2018bsc}. In particular,  emergent dark energy has drawn attention as it solves the tension between CMB and local measurements of $H_0$ while keeping the number of free parameters exactly as in flat $\LCDM$ cosmology. On the other hand, an
altogether different alternative explores modifications
to GR on  large scales.

As dark energy is a manifestation of cosmic accelerated expansion of the late universe,
a suitable weakening of gravity on cosmic distances at late times may
act in the same manner as dark energy, while recovering the successes of GR at early times and short distances; especially at  solar system (milliparsec) scales, or binary pulsar (microparsec) scales.
Such modifications may be referred to as infrared modifcations of gravity (IRMG).
Generally these modifications introduce new free parameters which may require further explanation and interpretation~\citep{Carroll:2006jn}. In this paper we test a particular form which does
not require any. It is achieved in terms an exponential IR modification to teleparallel equivalent
to GR:
$f(T)=T e^{\beta T_0/T}$, where $T$ is the teleparallel torsion scalar,
$T_0=-6 H_0^2$ and $\beta$ is a dimensionless parameter.
The theory has been previously introduced to provide a viable dynamical phase portrait compatible with the late transition from decelerated to accelerated expansion \citep{Awad:2017yod}. For $\beta > 0$, GR is recovered at early times and in strong gravity regimes, where $e^{\beta T_0/T} \to 1$. Consequently, the exponential IR $f(T)$ gravity is expected to be in agreement with the CMB observations and solar system tests.  It however modifies cosmic expansion on large distances at late time.

The model parameter $\beta$ turns out to be completely determined by the current values of the density parameters.  Therefore, unlike other viable $f(T)$ theories \citep{Nesseris:2013jea} or IRMG in general,
no extra parameters are introduced relative to standard cosmology.
The theory therefore does not embody, {\it a priori}, any additional freedom for fitting cosmological
data. It is thus statistically comparable, on equal footing, with $\LCDM$. In addition, this type of modification to GR acts effectively as a phantom DE without breaking the null energy condition \citep{Carroll:2003st,Carroll:2004hc,Ludwick:2017tox}. It can also completely resolve the $H_0$ tension between the CMB and local measurements without violating the age constraints, even if tension remains with BAO measurements \citep{El-Zant:2018bsc}.

In the present paper (hereafter paper I), we confront the exponential IR $f(T)$ theory with various cosmological data sets, in order to test its empirical viability.  The datasets used constrain the background expansion history, as well as the growth of linear perturbations
on scales well below the horizon, which turn are modified in a scale free manner.
The full linear perturbation analysis of the theory, with the full CMB powerspectra using Planck 2018 legacy, will be given in a companion paper~\citep{Hashim:2019b} (hereafter paper II). In Sec.~\ref{Sec:2}, we review $f(T)$ teleparallel gravity and its effect on the cosmological background evolution. In Sec.~\ref{Sec:3}, we discuss the particular $f(T)$ theory
studied here.  In Sec.~\ref{Sec:4}, we list and discuss the observational constraints  we consider,  and perform the joint likelihood analysis to obtain the best-fit values
for both $\LCDM$ and $\fTCDM$ models, with examining the empirical  viability of the proposed $f(T)$ theory in mind.
In Sec.~\ref{Sec:5}, we examine the viability of the $f(T)$ gravity in light of the obtained results. In Sec.~\ref{Sec:6} we summarise our conclusions and discuss prospects for future work.

\section{$\lowercase{f}(T)$ cosmology}\label{Sec:2}

We consider a $4$-dimensional $C^{\infty}$-manifold $(\mathcal{M},\,e_{a})$, where $e_{a}$ are four linear independent vector (tetrad, vierbein) fields defined on $\mathcal{M}$. The vierbein fields fulfil the conditions $e_{a}{^{\mu}}e^{a}{_{\nu}}=\delta^{\mu}_{\nu}$ and
$e_{a}{^{\mu}}e^{b}{_{\mu}}=\delta^{b}_{a}$, where the summation convention is assumed for both Latin (tangent $4$-spacetime coordinates), and Greek ($4$-spacetime coordinates)~\citep{Aldrovandi:2013wha,Krssak:2018ywd}.
The spacetime metric is related to the vierbein by
\begin{equation}
    g_{\mu \nu} \equiv \eta_{ab}e^{a}{_{\mu}}e^{b}{_{\nu}},
    \label{metric}
\end{equation}
where $\eta_{ab}$ is the tangent space Minkowski metric. Moreover,
one can straightforwardly construct the teleparalllel geometry by finding the nonsymmetric (Weitzenb\"{o}ck) linear connection $\Gamma^{\beta}{_{\mu\nu}}\equiv e_{a}{^{\beta}}\partial_{\nu}e^{a}{_{\mu}}=-e^{a}{_{\mu}}\partial_{\nu}e_{a}{^{\beta}}$.
Since $\Gamma^{\beta}{_{\mu\nu}}$ is nonsymmetric, it defines the torsion tensor
\begin{equation}
   T^\beta{_{\mu\nu}}\equiv{\Gamma^\beta}_{\nu\mu}-{\Gamma^\beta}_{\mu\nu}
   ={e_a}^\beta\left(\partial_\mu{e^a}_\nu
-\partial_\nu{e^a}_\mu\right),
\end{equation}
while its curvature vanishes identically. Thus, in this approach, gravity is encoded in terms of torsion instead of curvature. In this teleparallel geometry, one can define the torsion scalar,
\begin{equation}\label{Tor_sc}
    T=\frac{1}{4}T^{\alpha}{_{\mu\nu}}T_{\alpha}{^{\mu\nu}}
    +\frac{1}{2}T^{\alpha}{_{\mu\nu}}T^{\mu\nu}{_\alpha}
    -T^{\alpha}{_{\mu\alpha}}T^{\alpha\mu}{_\alpha},
\end{equation}
which is equivalent to the Ricci scalar $R$ up to a total derivative term. It therefore generates the same set of field equations as GR when it replaces $R$ in Einstein-Hilbert action. However theories representing
extensions of this teleparallel equivalent to GR,
known as $f(T)$ theories, differ from $f(R)$ in structure and consequences (see  the review \citep{Cai:2015emx} for details).

Cosmological models based on $f(T)$ gravity have been extensively explored~\citep{Camera:2013bwa,Nesseris:2013jea,Nunes:2018xbm,DAgostino:2018ngy,Abedi:2018lkr, Wang:2020zfv}.
However,  beyond background solutions and linear perturbations further work is required for understanding the foundational properties of $f(T)$ gravity. For, even the estimation of the number of degrees of freedom is different in different papers on Hamiltonian analysis  \citep{Li:2011rn,Ferraro:2018tpu,Blagojevic:2020dyq,Guzman:2019oth}.
Another important foundational point
relates to Lorentz invariance.. where more progress has been achieved.
Unlike in the case of $f(R)$,
the $f(T)$ field equations are not invariant under local Lorentz transformation in the
pure tetrad (trivial spin connection) formalism \citep{Li:2010cg,Sotiriou:2010mv}. However, a fully covariant version of the theory can be obtained by considering the spin connection contribution to the field equations \citep{Krssak:2015oua}. The misconception regarding their local Lorentz invariance in the $f(T)$ gravity has been also discussed in details in \citep{Krssak:2018ywd}.
In addition, with regards to the propagation of gravitational waves,
it has been shown that this corresponds to the speed of light, which makes such theories
compatible with the observation of GW170817 and its electromagnetic counterpart GRB170817A \citep{Cai:2018rzd}, see also \citep{Hohmann:2018jso}.

To evaluate the gravitational field produced by $f(T)$ gravity, we write the action
\begin{equation}
    \mathcal{S}=\frac{1}{2\kappa^2}\int d^{4}x\, |e|f(T)+\mathcal{S}_{M},
    \label{action}
\end{equation}
where $|e|=\sqrt{-g}=\det\left({e}_\mu{^a}\right)$, the constant $\kappa$ is related to the Newton's constant $G_N$ via $\kappa^2=8\pi G_N$ and  $\mathcal{S}_{M}$ is the action of the matter fields. The variation with respect to the vierbein gives rise to the field equations
\begin{equation}
    \frac{1}{\kappa^2_{\textrm eff}} \mathfrak{G}_{\mu\nu}= \mathfrak{T}^{(M)}_{\mu\nu}+\mathfrak{T}^{(DE)}_{\mu\nu},
\label{field_eqns}
\end{equation}
where $\kappa^2_{\textrm eff} = \kappa^2 / f_T$, and we take the perfect fluid approximation to describe the matter content
\begin{equation}
\mathfrak{T}^{(M)}_{\mu \nu}=\rho u_{\mu}u_{\nu}+p \left(u_{\mu}u_{\nu}+g_{\mu\nu}\right),
\end{equation}
where $\rho$, $p$ and $u^{\mu}$ are the density, pressure and 4-velocity unit vector of the fluid, respectively. This defines the ``\textit{geometrical}'' DE component via
\begin{equation}
\mathfrak{T}^{(DE)}_{\mu\nu}=\frac{1}{\kappa^2} \left(\frac{1}{2}g_{\mu\nu}\left(Tf_T-f\right)-f_{TT}S_{\nu\mu\rho}\nabla^{\rho}T\right).
\end{equation}

We assume the background geometry to be
flat Friedmann-Lema\^{\i}tre-Robertson-Walker (FLRW).
Hence, we take the Cartesian coordinate system ($t;x,y,z$) and the diagonal vierbein
\begin{equation}
   {e_{\mu}}^{a}=\textmd{diag}\left(1,a(t),a(t),a(t)\right),
\label{tetrad}
\end{equation}
where $a(t)$ is the scale factor of the universe. One can show that the above vierbein, via Eq. \eqref{metric}, generates the flat FLRW spacetime metric
\begin{equation}
   ds^2=dt^{2}-a(t)^{2}\delta_{ij} dx^{i} dx^{j},
\label{FRW-metric}
\end{equation}
where the Minkowskian signature is $\eta_{ab}=(+;-,-,-)$. We note that this choice of the vierbein in Eq.~\eqref{tetrad} is already in the proper form, since the associated spin connection is flat and subsequently leads to symmetric field equations for any $f(T)$ theory \citep{Krssak:2018ywd} (see also \citep{Ferraro:2011us,Krssak:2015oua,Cai:2015emx,Golovnev:2017dox}).
The diagonal vierbein Eq. \eqref{tetrad} directly relates the teleparallel torsion scalar (Eq.~\eqref{Tor_sc}) to Hubble rate as follows:
\begin{equation}
   T=-6H^2,
\label{TorHubble}
\end{equation}
where $H\equiv \dot{a}/a$ is Hubble parameter, and the ``dot'' denotes
differentiation with respect to the cosmic time $t$.
%
\subsection{$f(T)$ modification of the background dynamics}\label{Sec:2.1}
The field equations corresponding to the vierbein in Eq. \eqref{tetrad} give rise, respectively, to the Friedmann and Raychaudhuri equations
\begin{eqnarray}
&\frac{3}{\kappa^2} H^2 = \rho_m+\rho_r+\rho_T \equiv \rho_{\textrm eff},& \label{FR1T}\\
&-\frac{1}{\kappa^2}\left(3H^2+2\dot{H}\right) = p_r+p_T  \equiv p_{\textrm eff}.& \label{FR2T}
\end{eqnarray}
The torsion density and pressure, using \eqref{TorHubble}, in the above equations are given by
\begin{eqnarray}
&\rho_{T} = \frac{1}{2\kappa^2}\left(6H^2-f+Hf_H\right), &
\label{Tor-density}\\
&p_{T} = -\frac{1}{6\kappa^2}\dot{H}(12+f_{HH})-\rho_T, &
\label{Tor-press}
\end{eqnarray}
where $f= f(H)$, $f_H=\frac{df}{dH}$ and $f_{HH}=\frac{d^2 f}{dH^2}$.

Also, we write the continuity equations of the pressureless matter (baryon + cold dark matter), radiation and the torsion (respectively)
\begin{eqnarray}
&&\dot{\rho}_m+3H\rho_m = 0, \label{continuity_matter}\\
&&\dot{\rho}_r+4H\rho_r = 0, \label{continuity_radiation}\\
&&\dot{\rho}_T+3(1+w_T)H\rho_T = 0, \label{continuity_torsion}
\end{eqnarray}
where the torsion equation of state $w_T$ is given by
\begin{equation}
   w_{T}=\frac{p_{T}}{\rho_{T}}=-1+\frac{\left(f_{HH}+12\right) \left(f-H f_H \right)}{f_{HH} \left(f-6H^2-H f_H \right)}.
\label{torsion_EoS}
\end{equation}
It is useful to define the effective (total) equation-of-state parameter
\begin{equation}
  w_{\textrm eff}\equiv \frac{p_{\textrm eff}}{\rho_{\textrm eff}}=-1-\frac{2}{3}\frac{\dot{H}}{H^2},
\label{eff_EoS0}
\end{equation}
which can be related to the deceleration parameter $q$ by the following expression
\begin{equation}
   q\equiv -1-\frac{\dot{H}}{H^2}=\frac{1}{2}\left(1+3 w_{\textrm eff}\right).
\label{deceleration}
\end{equation}
A nice feature of the $f(T)$ field equations is that they are of second order, unlike those
associated with other gravitational theories; e.g., the $f(R)$ field equations, which are fourth order. Furthermore, this family of governing equation of any $f(T)$ theory reduces to a one-dimensional autonomous system, similar to GR. This makes the $f(T)$ gravity a natural generalization of GR, whereas the governing equation is given by \citep{Awad:2017yod}
\begin{equation}
    \dot{H}=3(1+w)\frac{f-H f_{H}}{f_{HH}}= \mathcal{F}(H).
\label{phase_portrait_fT}
\end{equation}
This dynamical view enables the succinct visualizion of the global
dynamics of the system
through the corresponding phase portrait in the $H,\, \dot{H}$ phase--space.
%
\subsection{$f(T)$ modification of growth function}\label{Sec:2.2}
In the framework of  $f(T)$ modified gravity, at  Newtonian sub-horizon scales, one can expect that the dark energy is smooth and consider linear perturbations only on the matter sector, whereas changes in the evolution of the growth of structure is determined by modified background expansion effects on the growth rate. In particular, the Hubble expansion rate $H$ and the effective Newton’s gravitational constant $\kappa^2_{eff}=\kappa^2/f_T$; c.f. \citep{Nesseris:2013jea, Golovnev:2018wbh,Anagnostopoulos:2019miu} and discussion below.

In modified gravity, generic modification of the dynamics of the linear matter perturbation at subhorizon scales can be represented via the matter continuity equation \citep{Amendola:2016saw}
\begin{equation}
    \Delta''_m + \left(2 + \frac{H'}{H} \right) \Delta'_m - \frac{3}{2} \Omega_m(a)\frac{Q(a,k)}{\eta(a,k)}\Delta_m = 0,
    \label{eq:growthrate}
\end{equation}
for the comoving matter density perturbation $\Delta_m$ where prime denotes the derivative $d/d\ln a$ and $\Omega_m(a)=(8\pi G_N/3H^2)\rho_m$. In $f(T)$ gravity, in particular on the sub-Hubble scale, the parameters are $Q(a,k)= Q(a)=1/f_T$ (strength of gravity modified by the factor of $f_T$) and $\eta(a,k)\to 1$ (no gravitational slip in the subhorizon limit, even though it is an important contribution to the superhorizon regime). Of course in case of $f(T)=T$ the standard GR is recovered.

In the present context,
the evolution of the comoving density contrast $\Delta_m$ in $f(T)$
is scale independent as in GR but driven by the product $\Omega_m(a) Q(a)=\Omega_m(a)/f_T$. Therefore, in this context,  $f(T)$ gravity modifies the growth of structure in a way very similar to GR. This is in contrast to $f(R)$ gravity which includes scale dependence effects.
The  formulation here however is valid solely at sub-horizon scale $k\gg aH$, otherwise the equation for growth has a complicated form \citep{Li:2011cg,Golovnev:2018wbh}, which may lead to a large deviation from $\LCDM$ in the matter power spectra on large scales. This will be revisited in paper II \citep{Hashim:2019b}, when the full perturbation analysis is adopted.

As linear matter perturbations in the Newtonian limit are sensitive to the background modifications through $H(a)$ and the effective Newton's constant $\kappa^2/f_T$, one can include growth rate observations in the viability test of the exponential IR $f(T)$ gravity at the background level.
We define the growth factor of the matter density contrast
\begin{equation}
\label{eq:growthfactor}
    G=\frac{\Delta_m(a,k)}{\Delta_{m,0}(1,k)},
\end{equation}
where $\Delta_{m,0}(1,k)$ is the comoving density contrast current value. In practice, one considers the product ${f}\sigma_8$ to test the viability of the model with the red-shift space distortion observations, where the cosmological growth rate is given by
\begin{equation}
\label{eq:growth_fn}
    {f}=\frac{d\ln G}{d\ln a},
\end{equation}
and $\sigma_8$ is the standard deviation of the overdensity $\delta_m$ measured in spheres of radius $8\, h^{-1} Mpc$. By solving Eq. \eqref{eq:growthrate} with \eqref{eq:growthfactor} and \eqref{eq:growth_fn}, one can confront the theory with the RSD observations (see Sec. \ref{sec:growthrate}).

\section{The Exponential infrared $f(T)$ Gravity}\label{Sec:3}

\subsection{Motivation}

Several $f(T)$ theories have been proposed in the literature with the aim of realizing late accelerated expansion. Such theories are generally
characterized by one or two model parameters~\citep{Bengochea:2008gz,Linder:2010py,Bamba:2010wb}. Under particular choices of the parameters,  models that have been hitherto shown viable essentially reduce to $\LCDM$ cosmology but include an extra parameter relative to that standard model. This makes the latter preferable given similar fits to the data.
For example in the case of the power-law theory $f(T)=T+\alpha (-T)^{n}$ which invokes a new independent parameter $n$, it has been shown that the parameter $n$ is almost null when $H_0$+SNIa+BAO+CMB dataset is used. This effectively Other $f(T)$ theories which do not cover $\LCDM$ as a particular case have been shown to be non-viable \citep{Nesseris:2013jea,Xu:2018npu}.

In light of this, it becomes apparent that finding MG theories which    exports $\alpha$ into Friedmann equation as a cosmological constant.
provide a viable fundamental alternative
to $\LCDM$ without introducing new parameters, is non-trivial.
In the rest of this section, we describe one such model and outline
its properties.

\subsection{The model}

\subsubsection{Basic form}

By examining the generic phase portrait patterns of viable models, an $f(T)$ theory has been proposed to produce late accelerated expansion \citep{Awad:2017yod}
\begin{equation}
f(T)=T e^{\beta (T_{0}/T)},
\label{exp-IR}
\end{equation}
where $T_0 = -6H_0^2$ and $\beta$ is a dimensionless parameter.

The GR limit of this model is recovered by setting $\beta=0$.
Likewise, in the early universe and in strong gravity regimes, $T\gg \beta T_0$, the exponential correction factor goes to unity. Therefore the GR limit is recovered and no conflicts with CMB or solar system observations are expected. We now show that $\beta$ is effectively
entirely fixed by the matter density.

\subsubsection{No extra parameters}

Following \citep{Nesseris:2013jea}, the modified Friedmann equation can be rewritten as
\begin{equation}
E^2(z)=\Omega_{m}(1+z)^3+\Omega_{r}(1+z)^4+\Omega_{T} y(z). \label{FR-E}
\end{equation}
Here $E(z)=H(z)/H_0$, $\Omega_{i}$ is current value of density parameter (the subscript $i$ denotes $m$, $r$ and $T$ for matter, radiation and torsion), $\Omega_T=1-\Omega_m-\Omega_r$, and the distortion function $y(z)$ is given by
\begin{equation}
    y(z)=\frac{6H^2-f(H)+Hf_H}{6\Omega_{T}H_0^2},
    \label{y(z)}
\end{equation}
where we adopted $T-H$ relation, namely Eq.~\eqref{TorHubble}. From Eq.~\eqref{exp-IR}, the distortion function becomes
\begin{equation}
y(z)=\Omega^{-1}_{T}\left(E^2-(E^2-2\beta)e^{\frac{\beta}{E^2}}\right),
\label{y1(z)}
\end{equation}
and consequently the Friedmann equation (Eq.~\eqref{FR-E}) reads
\begin{equation}
\left(E^2-2\beta\right) e^{\frac{\beta}{E^2}}=\Omega_{m}(1+z)^3+\Omega_{r}(1+z)^4. \label{FR-E-exp-IR}
\end{equation}
At the present epoch, i.e $z=0$ and $E=1$, the $\beta$-parameter can be expressed in terms of the current values of the density parameters as
\begin{equation}
    \beta=\frac{1}{2}+\mathcal{W}\left(\frac{\Omega_{m}+\Omega_{r}}
    {-2e^{\frac{1}{2}}}\right), \label{beta}
\end{equation}
where $\mathcal{W}(x)$ is the Lambert $\mathcal{W}$ function\footnote{Defined via $x=\mathcal{W}(x) \, \exp{\mathcal{W}(x)}$.}. At present, the radiation density parameter $\Omega_{r}=\Omega_{\gamma}\left[1+\frac{7}{8}\left(\frac{4}{11}\right)^{4/3} N_{\textrm eff}\right]$, whereas the photon density parameter $\Omega_{\gamma}=2.4728 \times 10^{-5} h^{-2}$, and the effective number of neutrino species as in standard model $N_{\textrm eff}=3.046$.

Thus the advantage of this model is that it does not introduce any new parameters in the Friedmann equation, Eq. \eqref{FR-E-exp-IR}, other than those in $\LCDM$, i.e $\left\{\Omega_{m},\, H_0\right\}$.

\subsubsection{Basic properties}

For larger Hubble values, the model reduces to GR. In contrast, in the small Hubble regime one expect deviations from the GR limit on large scales. This  gives rise to accelerated expansion that does not necessarily
correspond to that induced by a cosmological constant, while keeping GR's successes at solar system and astrophysical scales (the detailed scale dependence of this weakening is discussed in the linear regime in \citep{Hashim:2019b}).

\begin{figure}
    \centering
    \includegraphics[scale=0.6]{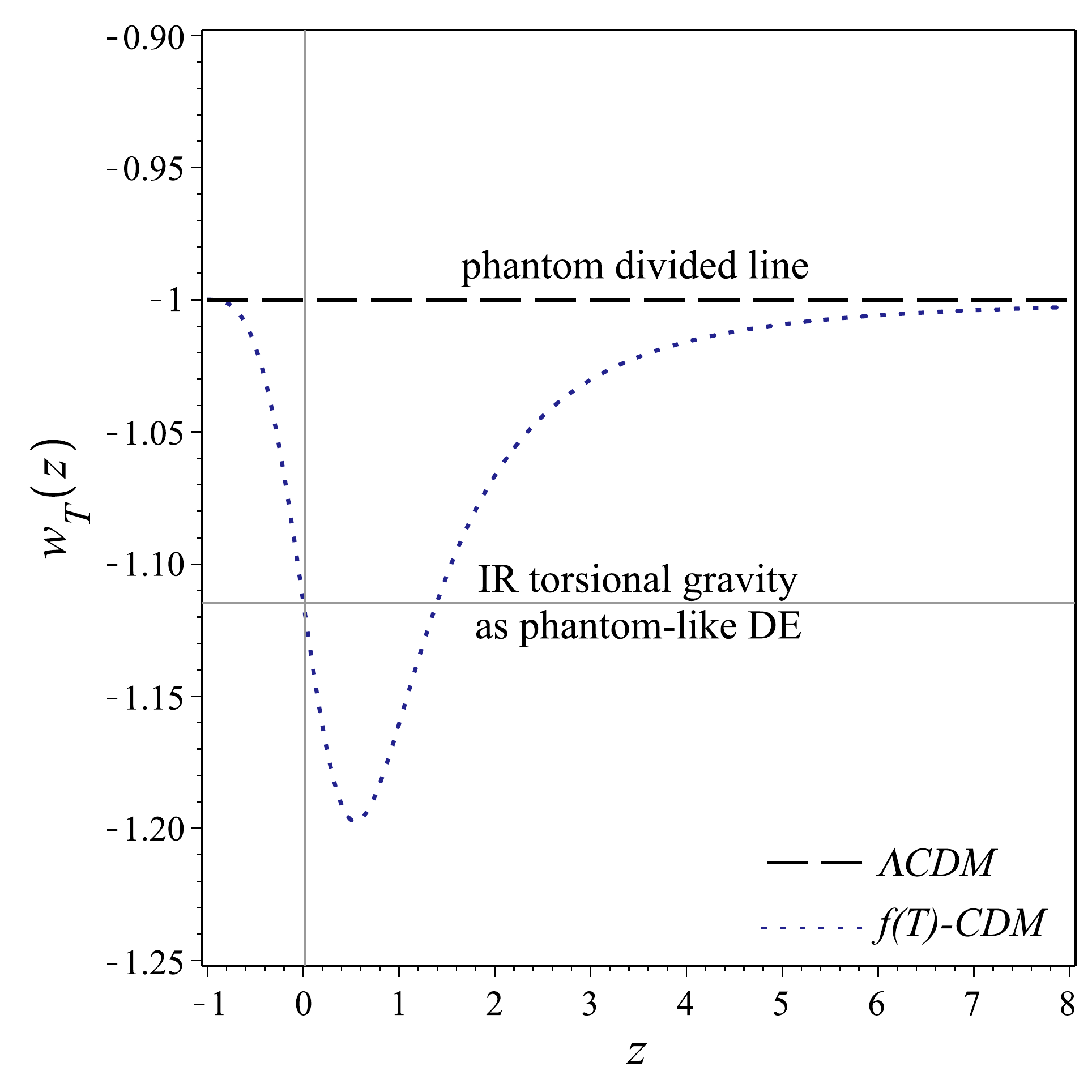}
    \caption{The evolution of torsional gravity counterpart. It illustrates the effective dark energy role and
    phantom-like nature of the IR gravity corrections.}
    \label{fig:Tor_EoS}
\end{figure}

At redshift $z\to -1$, one finds $E^2 \to 2\beta$. This is in fact a future de Sitter fixed point but pushed up to $t\to \infty$.
Recalling Eq.~\eqref{torsion_EoS}, one also finds that the torsional gravity counterpart, $w_T$, evolves as a phantom-like DE, as probed in Fig. \ref{fig:Tor_EoS}. As is well known, the infrared (IR) correction of gravity produces an apparent phantom dark energy $w<-1$ without violating the null energy condition \citep{Carroll:2004hc,Ludwick:2017tox}.

In Fig.~\ref{fig:Tor_EoS} we show the general behavior of torsion acting as DE.
At large $z$, we have $w_T(z)\to -1$,  as with a cosmological constant. Nevertheless,
$\rho_T=-p_T\to 0^+$, unlike a cosmological constant,
which has fixed density and pressure at all time. At low redshifts $z\sim 8$, the torsional counterpart evolves as  phantom DE. At present $w_T \thickapprox -1.12$, while it is evolving towards pure de Sitter spacetime with $w_T\to -1$ as $z\to -1$ (i.e $t\to \infty$).

\section{Observational Constraints and datasets}\label{Sec:4}
We employ different datasets to constrain the exponential IR $f(T)$ gravity, testing its viability as a model of late-time cosmic acceleration. The same analysis  is applied to  $\LCDM$ for comparison. Remarkably, in both cases, the background evolution parameters
also effectively fix the parameters determining the height of the CMB peaks, while only their angular location is explicitly used here, cf. Section \ref{Sec:5.4}.

In the following subsections we give a brief description of those different datasets and the methodology used in the present analysis.
\subsection{Astronomical datasets}\label{Sec:4.1}
\subsubsection{Supernovae type Ia}\label{Sec:4.1.1}
Type Ia supernovae as standard candles have been crucial to cosmology since leading
to the discovery of cosmic acceleration in the late 1990's~\citep{Riess:1998cb,Perlmutter:1998np}.
In comparison with BAO and CMB data, SNIa data is however
statistically less potent in constraining $\LCDM$ model in general.
They remain nevertheless
essential for testing background cosmological
evolution models at low redshifts, which is
the main concern of our present analysis.

The SNIa distance modulus $\mu$ is related to the luminosity distance $D_L$ via the relation
\begin{equation}
    \mu(z) \equiv m(z) - M = 5 \log_{10}(D_L) + 25,
    \label{Eq:distmodulus}
\end{equation}
where $m$ is the apparent magnitude and $M$ is the absolute magnitude. The distance $D_L$, in Mpc, is given by
\begin{equation}
    D_L = \frac{(1+z)}{H_0} \int_0^z \frac{d\acute{z}}{E(\acute{z})}, \label{Eq:lumdist}
\end{equation}
for a flat FLRW model (i.e. $\Omega_K = 0$). As in Eq. \eqref{Eq:lumdist}, the luminosity distance is solely determined  by the modified Friedman equation, Eq.~\eqref{FR-E-exp-IR}. To get the background dynamics of the $\fTCDM$ model, we numerically solve Eq. \eqref{FR-E-exp-IR} by iteration. Pantheon SNIa observed distance modulae are then calculated using the B-Band apparent and absolute magnitudes. The absolute magnitude $M$ is almost constant for all supernovae and is taken as an inference parameter.

We use the Pantheon sample from \citep{Scolnic:2017caz},
with $276$ additional supernovae to the Joint Light-curve Analysis (JLA) sample \citep{Betoule:2012an,Betoule:2014frx,Mosher:2014gyd} from the Pan-STARRS1 Medium Deep Survey,  plus low-redshift and Hubble space telescope (HST) samples. This comes to a
total of $1048$ supernovae spanning redshift range $0.01 < z < 2.3$.

We note that from Eqs. \eqref{Eq:distmodulus} and \eqref{Eq:lumdist}, the SNIa distance modulus only constrains parameters in the function $E(z)$. This is known as distance-redshift degeneracy \citep{Amendola:2015ksp}; the absolute magnitude $M$ is degenerate with the Hubble constant $H_0$.

\subsubsection{$H_0$ measurements}\label{Sec:4.1.2}
We take, as a prior, the value of the Hubble constant recently measured by a distance ladder method, using 75 Milky Way Cepheids with HST + Gaia EDR3 parallaxes, combined with best complementary sources of Cepheid calibration: $H_0=73.2 \pm 1.3$ km/s/Mpc with 1.8\% precision \citep[R20]{Riess:2020fzl}.
\subsubsection{Baryon acoustic oscillations}\label{Sec:4.1.3}
We consider BAO radial measurements of $H(z)r_{drag}$ along the line of sight, as well as the BAO transverse measurements of $D_V(z)/r_{drag}$ perpendicular to the line of sight. Here $r_{drag}$ is the comoving sound horizon at the end of the baryon drag, and $D_V$ is a combination of the comoving angular distance $D_M(z)=D_L/(1+z)$ and the Hubble parameter $H(z)$ given by
\begin{equation}
    D_{V}(z) = \left[D^2_M(z)\frac{cz}{H(z)}\right]^{\frac{1}{3}}.
\label{eq:DV}
\end{equation}

We use the high precision measurements of the latest BOSS data release 12 (BOSS DR12) \citep{Alam:2016hwk}, which summarized "consensus" results on BAOs (first reported in \citep{Ross:2016gvb,Beutler:2016ixs} and \citep{Vargas-Magana:2016imr}) at effective redshift bins $z_{\textrm eff} = 0.38, 0.51$ and $0.61$. In addition, we consider the two measurements of $D_V/r_{drag}$ at low redshifts $z_{\textrm eff}=0.106$ and $z_{\textrm eff}=0.15$ by the 6dFGS \citep{2011MNRAS.416.3017B} and SDSS-MGS \citep{Ross:2014qpa}, respectively. We also use the WiggleZ redshift survey reconstructed measurements \citep{Kazin:2014qga}, as well as the recent BAO measurement by eBOSS DR16, using multi-tracers in configuration space, at $z = 0.77$ for $D_H \equiv c/H(z) = 19.65 \pm 0.54 \times r_{drag}$  and $D_M/r_{drag} = 18.93 \pm 0.37$ \citep{Wang:2020tje}. We note that, at low redshift, the combined BAO likelihood is dominated by the high precision measurements of BOSS DR12.

\subsubsection{Redshift-space distortion}\label{Sec:4.1.4}
The peculiar motions of galaxies, relative to the Hubble flow,
introduce anisotropies in the galaxy clustering observed in reshift surveys. This phenomenon is known as redshift-space distortions (RSD). The measurements of RSD could constrain the amplitude of the matter power spectrum, and in turn the structure growth rate \citep{Percival:2008sh}. Usually, measurements of RSD are given in terms of ${f}\sigma_8$, where the growth rate ${f}$ is as given via \eqref{eq:growth_fn}. For $\LCDM$, the growth rate is approximated by the parametrization ${f} \sim \Omega^{0.55}_m(z)$. For the $\fTCDM$ model, we numerically solve Eq.~\eqref{eq:growthrate} for the growth function $\Delta_m$.

We use RSD measurements of ${f} \sigma_8$ from BOSS DR12 results \citep{Ross:2016gvb}, together with WiggleZ \citep{2012MNRAS.425..405B}, eBOSS DR16 \citep{Wang:2020tje}, SDSS MGS \citep{Howlett:2014opa}, 6dFGRS \citep{2012MNRAS.423.3430B}, and the growth rate constraint by \citep{Said:2020epb} (obtained by comparing observed fundamental plane peculiar velocities in 6dFGS with predicted velocities and densities from the 2M++ redshift survey).
\subsection{Cosmological constraints and allowed parameter space}\label{Sec:4.2}
\subsubsection{BBN constraint on Baryon density}\label{Sec:4.2.1}
We use the conservative prior on the baryon density $\omega_b=\Omega_b h^2 = 0.0222 \pm 0.0005$ (68\% CL), as calculated by Planck 2018  \citep{Aghanim:2018eyx}, and found to be compatible with the three BBN calculation pipelines based on the deuterium abundance measurement \citep{Cooke:2017cwo}.
\subsubsection{CMB constraints}\label{Sec:4.2.2}
We add a conservative CMB-BAO measurement
of the angular acoustic scale at decoupling $\theta_{\textrm s}=\theta(z_{\textrm s})$, where $z_{\textrm s}$ defines the redshift at which the optical depth equals unity; i.e., $\tau(z_{\textrm s})=1$. We use the constraints on the base parameters obtained from Planck 18 (TT,TE,EE+lowE+lensing) dataset \citep{Aghanim:2018eyx}, to obtain the value of 100$\theta_{\textrm s} = 1.04190 \pm 0.00030$ using the  {\verb"CLASS"} code. Since this parameter is measured with a precision of sub-percent level, the procedure allows for a tight constraint on the parameter space, comparable to those obtained from the full CMB dataset.

We note that the value derived for $\theta_{\textrm s}$ using the {\verb"CLASS"} code (in the current analysis) agrees with the $\theta_{\textrm MC}$ presented in Planck results (which is derived using {\verb"CAMB"} and {\verb"CosmoMC"} codes) within 1$\sigma$. It is also worth  mentioning that  $\theta_s$ is the actual angular scale of the sound horizon at decoupling, obtained by fully integrating over the sound speed and then searching numerically for the time of decoupling (defined as the maximum of the visibility function). On the other hand, $\theta_{\textrm MC}$ is an approximation based on a model-dependent analytical fits instead of the full integral \citep{2011JCAP...07..034B}.

Finally, we fix two parameters as measured by CMB Planck 2018~\citep{Aghanim:2018eyx}, namely the optical depth at reionization $\tau(z_{re})= 0.0544 \pm 0.0073$ and the spectral index $n_s= 0.9649 \pm 0.0042$.
\subsubsection{Parameter space}\label{Sec:4.2.3}
For both the $\LCDM$ and $\fTCDM$ models, we take the Base dataset as SNIa + $H_0$ + BBN + BAO, which allows to fix three parameters, namely the Hubble parameter and the baryon and CDM densities, such that $\left\{H_0,~ \Omega_b,~\Omega_c\right\}$. By adding the RSD data, we can fix one more parameter, that is the amplitude of the growth of structure $\left\{\sigma_8 \right\}$. In addition, the inclusion of the CMB $\theta_{\textrm s}$ allows for better constraints on the full parameter space
\begin{equation}\label{parameter_space}
    \mathcal{P}=\left\{H_0,~ \Omega_b,~\Omega_c;~\sigma_8 \right\}.
\end{equation}

We also derive additional parameters, namely $\left\{\Omega_m,~\hat{\theta}_s,~r_{drag},~z_{re},~S_8\right\}$. Here $\hat{\theta}_s$ is the angular size of the sound horizon at recombination
, $z_{re}$ is the reionization redshift and $S_8=\sigma_8\sqrt{\Omega_m/0.3}$. In
addition, to the absolute magnitude of supernova, $M$,
taken as an inference parameter.

We use the \verb"CLASS" code \citep{2011JCAP...07..034B}, together with \verb"Monte Python" \citep{2013JCAP...02..001A}, after proper modifications (for the computation of $\fTCDM$ background dynamics) to run Monte Carlo Markov Chain (MCMC) analysis. In order to analyze the resulting MCMC chains and obtain the contour plots of the different model parameters, we use of \verb"GetDist" python package \citep{Lewis:2019xzd}.
\section{Results}\label{Sec:5}
In this section we test the viability of the $\fTCDM$ theory with a full likelihood analysis and compare it to $\LCDM$. We also examine the consistency of the obtained results in light of the different datasets listed and described above. We point out the recent tensions between different datasets and discuss prospects for resolution within the exponential IR $f(T)$ theory.
\subsection{Viability of the $\fTCDM$ model}\label{Sec:5.1}
As  the current $f(T)$ theory has the same number of free parameters as $\LCDM$ model, a direct statistical comparison using a $\chi^2$ is straightforward.
\begin{table*}
\caption{$68\%$ parameter intervals for $\LCDM$ and $\fTCDM$ models from SNIa, $H_0$, BAO, BBN, RSD and CMB $\theta_{ s}$ measurements grouped in there datasets. The minimum value of $\chi^2$ for each model is given in the last row. Here ``\textit{Base}" represents SNIa, $H_0$, BBN and BAO joint dataset.}\label{Table:datasets}
\centering
\resizebox{\columnwidth}{!}{%
\begin{tabular}{ l c c c c c c}
\hline\hline
\multirow{3}{*}{Parameter}& \multicolumn{3}{c}{$\LCDM$} & \multicolumn{3}{c}{$\fTCDM$} \\
\cline{2-7}
& Base & Base+RSD & Base+RSD+CMB $\theta_{\textrm s}$ & Base &Base+RSD & Base+RSD+CMB $\theta_{\textrm s}$ \\
& $68\%$ limits & $68\%$ limits & $68\%$ limits & $68\%$ limits & $68\%$ limits & $68\%$ limits \\
\hline
{$H_0          \dotfill    $} & $ 68.3^{+1.0}_{-1.3} $ & $ 68.4\pm 1.0 $ & $ 68.30\pm 0.77 $ & $ 70.7\pm 1.3 $ & $ 70.6^{+1.3}_{-1.2} $ & $ 70.52\pm 0.71 $ \\[5pt]
{$\Omega_{b}   \dotfill $} & $ 0.0477\pm 0.0017 $ & $ 0.0477^{+0.0015}_{-0.0017} $ & $ 0.0478\pm 0.0011 $ & $ 0.0438^{+0.0014}_{-0.0017} $ & $ 0.0440^{+0.0013}_{-0.0016} $ & $ 0.04485\pm 0.00090 $ \\[5pt]
{$\Omega_{c} \dotfill $} &$ 0.256^{+0.017}_{-0.020} $ & $ 0.255\pm 0.017 $ & $ 0.2516\pm 0.0086 $ & $ 0.272\pm 0.018 $ & $ 0.270\pm 0.018 $ & $ 0.2486\pm 0.0072 $ \\[5pt]
{$\sigma_8      \dotfill  $} &-- & $ 0.782\pm 0.024 $ & $ 0.781\pm 0.025 $ & -- & $ 0.766\pm 0.033 $ & $ 0.781^{+0.035}_{-0.040} $ \\[5pt]
\hline
{$\Omega_{m}   \dotfill $}& $ 0.305^{+0.016}_{-0.018} $ & $ 0.304\pm 0.016 $ & $ 0.3008\pm 0.0092 $ & $ 0.317\pm 0.017 $ & $ 0.315\pm 0.017 $ & $ 0.2947\pm 0.0077 $ \\[5pt]
{$100\hat{\theta}_{\textrm s}  \dotfill$} & $ 1.044^{+0.014}_{-0.016} $ & $ 1.044^{+0.015}_{-0.013} $ & $ 1.04189^{+0.00030}_{-0.00033} $ & $ 1.055^{+0.017}_{-0.015} $ & $ 1.054^{+0.016}_{-0.014} $ & $ 1.04192\pm 0.00030 $ \\[5pt]
{ $r_{ drag} \dotfill$} & $  147.413 ^{+ 7.594 }_{- 7.626 }$ &$  147.377 ^{+ 7.256 }_{- 7.387 }$ & $  147.821 ^{+ 4.322 }_{- 4.981 }$ &$  143.611 ^{+ 11.349 }_{- 8.026 }$ & $  143.913 ^{+ 10.767 }_{- 7.88 }$ &$  146.189 ^{+ 5.496 }_{- 4.385 }$  \\[5pt]
{$z_{ re}   \dotfill $}& $ 7.51^{+0.22}_{-0.25} $ & $ 7.49\pm 0.23 $ & $ 7.46\pm 0.13 $ & $ 7.84\pm 0.24 $ & $ 7.81^{+0.26}_{-0.22} $ & $ 7.53\pm 0.12 $ \\[5pt]
{$S_{8} \dotfill$}& -- & $ 0.786\pm 0.025 $ & $ 0.782\pm 0.025 $ & -- & $ 0.785\pm 0.035 $ & $ 0.774^{+0.035}_{-0.041} $ \\[5pt]
\hline
{$M              $}& $ -19.401^{+0.038}_{-0.046} $ & $ -19.399\pm 0.037 $ & $ -19.402\pm 0.021 $ & $ -19.352\pm 0.047 $ & $ -19.355^{+0.047}_{-0.042} $ & $ -19.365\pm 0.019 $ \\[5pt]
\hline
{ $\chi^2_{ min}$} & $ 519.411 $ & $ 523.893 $ & $ 525.862 $ & $ 516.441 $ & $ 518.091 $ & $ 524.465 $ \\
\hline\hline
\end{tabular}
}
\end{table*}
\begin{figure*}
    \centering
    \includegraphics[width=\textwidth]{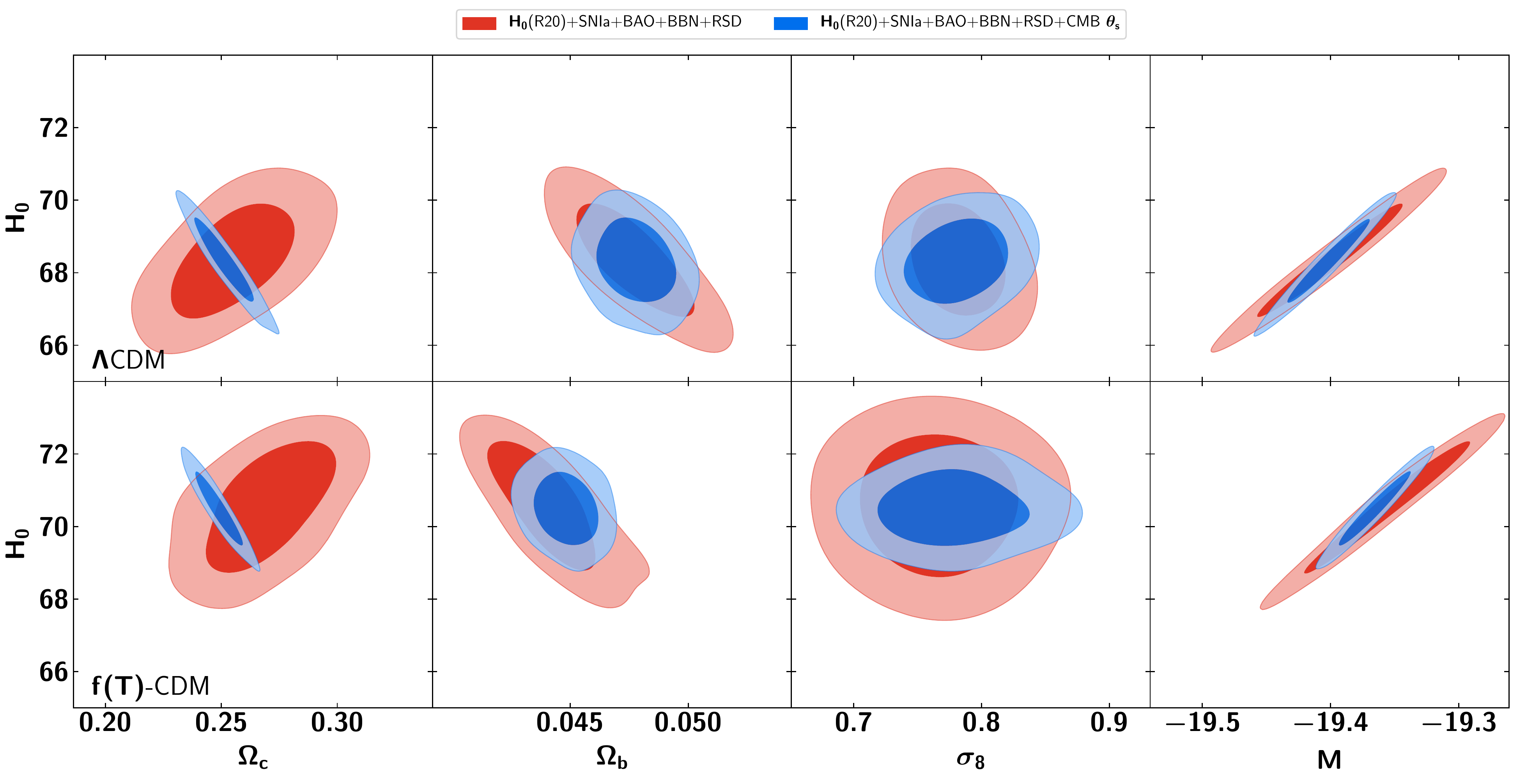}
    \caption{A compilation of $\sgmI$ and $\sgmII$ contour plots for the CDM and baryon density parameters, the nonlinear scale fluctuations $\sigma_8$,
    and the supernovae magnitude $M$ (as inference parameter), against $H_0$. This is shown for both $\LCDM$, in the upper panel, and $\fTCDM$ model, in the lower panel. Red contours represent the joint likelihood analysis for SNIa, H(z), BAO, BBN and RSD datasets, while blue contours include in addition the BAO measurement of the CMB (namely the angular acoustic scale $\theta_{\textrm s}$ measurement).}
    \label{Fig:mdlcomb}
\end{figure*}

In Table~\ref{Table:datasets}, we list the best-fit values up to 68\% CL, for both $\LCDM$ and the exponential IR $f(T)$ theory, using different combinations of cosmological datasets; such as Base ($\equiv$ SNIa+$H_0$+BBN+BAO), Base+RSD and Base+RSD+CMB $\theta_{\textrm s}$. As  can be seen, the minimum Chi-squares ($\chi^2_{min}$) for both models are comparable, with slight preference for $f(T)$ in all dataset combinations, which confirms the viability of the $\fTCDM$ model. We note that by utilizing the concise CMB measurement of the angular scale of the sound horizon, $\theta_{\textrm s}$, we obtain more constrained values of the inferred parameters, while keeping the agreement of the two models within 1$\sigma$.

In Fig.~\ref{Fig:mdlcomb}, we plot the 2D joint contours of the model parameters $\left\{\Omega_c,~\Omega_b,~\sigma_8\right\}$ versus $H_0$, for $\LCDM$ and $\fTCDM$ scenarios at 68\% and 95\% confidence level (CL). This was done using the full likelihood analysis for the full set of parameters, including the inferred 'nuisance' parameter $M$, for two main datasets (with/without CMB $\theta_{\textrm s}$ constraint). As is apparent, with the inclusion of the  Planck constraint on the CMB $\theta_{s}$, both $\LCDM$ and $\fTCDM$ remain in agreement, while the latter gives higher $H_0$ value compared to $\LCDM$, indicating a partial solution of the associated tension.

\subsection{Consistency in light of various observations}\label{Sec:5.2}
We examine the consistency of the obtained results from the joint MCMC likelihood  with individual observational datasets; namely how they fare separately with SNIa, BAO and RSD. We use the best fit values, in particular the full dataset combination Base+RSD+CMB $\theta_{\textrm s}$, as provided by Table \ref{Table:datasets} for both the $\LCDM$ and the exponential IR $f(T)$ models.
\begin{figure}[h!]
    \centering
    \includegraphics[scale=0.35]{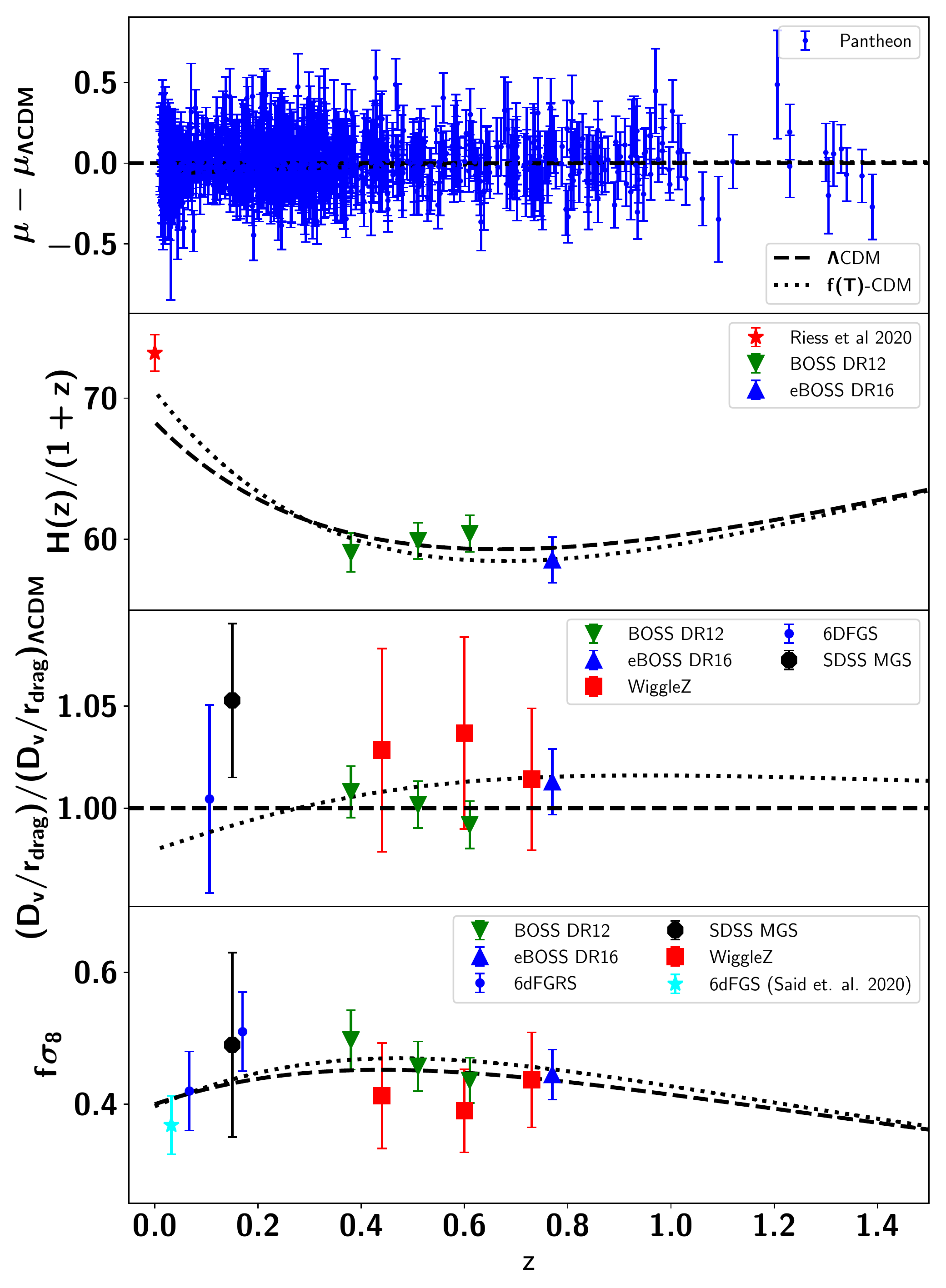}
    \caption{Various datasets are compared with theoretical predictions of $\LCDM$ and $\fTCDM$ models, as given in Table \ref{Table:datasets}. {Upper panel:} Residual of SNIa distance modulus measurements (blue dots) of Pantheon data sample from $\LCDM$ model. {Second panel:} Hubble parameter $H_0$ from R20 local measurements (red star), BOSS DR12 BAO radial distance measurements (green down triangles) and eBOSS DR12 (blue up triangle). { Third panel:} the ratio of $D_V$ (Eq. \ref{eq:DV}) over the comoving sound horizon at the baryon drag $r_{drag}$, with respect to $\LCDM$ model from BOSS DR12 (green down triangles), WiggleZ (red squares), eBOSS DR12 (blue up triangle), low redshift 6DFGS (blue dot) and SDSS MGS (black pentagon) measurements.  {Bottom panel:} $f\sigma_8$ measurements from the same dataset as in the third panel in addition to 6dFGRS (blue dots) and the recent 6dFGS  measurement (cyan star) \citep{Said:2020epb} .}
    \label{Fig:datasets}
\end{figure}

In the upper panel of Fig.~\ref{Fig:datasets}, Pantheon data are compared with $\LCDM$ and $\fTCDM$ models with best-fit parameters values. Both are in a good agreement with SNIa data. We note that the $\LCDM$ and $\fTCDM$ models give, respectively, absolute magnitudes $M= -19.402 \pm 0.021$ and $M= -19.365 \pm 0.019$ which are close to the true absolute magnitude of SNIa, $M=-19$.

In the second panel of Fig.~\ref{Fig:datasets}, we show how well $\LCDM$ model fits $H(z)$ BAO measurements, as provided by the precise constraints of BOSS DR12 and the recent eBOSS DR16 observations over redshift range $0.3 < z < 1$.  However, it fails to reach local measurement of $H_0$ value at $z = 0.0$ \citep{Riess:2018byc}. On the other hand, $\fTCDM$ tends to reach higher $H_0$ value in better agreement with the local $H_0$ measurement while keeping the good fit with the BAO $H(z)$ measurements.

In the third panel of Fig.~\ref{Fig:datasets}, we plot various BAO data used in this analysis in comparison with the theoretical prediction from $\fTCDM$ model. We show the distance of the acoustic-scale ratio $D_V/r_{drag}$ at several effective redshifts (as given in the figure), divided by the acoustic-scale ratio in the $\LCDM$ model. Both $\LCDM$ and $\fTCDM$ (with best fit parameters) seem to agree very well with BAO measurements. We note that the exponential IR $f(T)$ gravity entails a relatively mild phantom regime later in cosmic expansion history relative to the power law
models discussed in \citep{El-Zant:2018bsc}. Additionally, the current treatment only partially alleviates the $H_0$ tension. In this context, the stark inconsistencies  with BAO distances found in the aforereferenced work are avoided.

In the bottom panel of Fig.~\ref{Fig:datasets}, we use the best fit values of Table~\ref{Table:datasets} to plot the theoretical predictions of the rate of the growth of structure diagnostic ${f}\sigma_8$ for $\LCDM$ and $\fTCDM$ models. Both models seem to agree perfectly with the RSD dataset.

Another consistency test of the exponential IR $f(T)$ gravity is the age of the universe as predicted by the theory. According to the full likelihood results, as given in Table \ref{Table:datasets}, the age of the universe is $\sim 13.76$ Gyr, which is not in conflict with any of the known astrophysical observations so far \citep{Valcin:2020vav,2018ApJ...867...98S,Jimenez:2019onw}.

We conclude that the exponential IR $f(T)$ theory is statistically similar to $\LCDM$, since both have the same number of free parameters, and the best-fit $\chi^2$ results are almost the same
for the different dataset combinations in Table \ref{Table:datasets}. The $f(T)$ theory however shows some deviations at low redshifts as clear from Fig.~\ref{Fig:datasets} with $H(z=0)$ and $f\sigma_8$. We focus on the tensions related to these quantities in the following.

\subsection{The Hubble constant}\label{Sec:5.3}
As is already clear from Table \ref{Table:datasets}, the best fit values in parameter space
for $\LCDM$ and the $\fTCDM$ model are recognizably different when the CMB $\theta_{\textrm s}$ is absent from the joint MCMC analysis (namely the Base and the Base+RSD combined data). However, the inclusion of the CMB $\theta_{\textrm s}$ results in consistent values for the two models.
This is understandable, as the angular acoustic scale is observationally pinpointed with $\sim 0.03\%$ precision.

\begin{figure}
    \centering
    \includegraphics[scale=0.4]{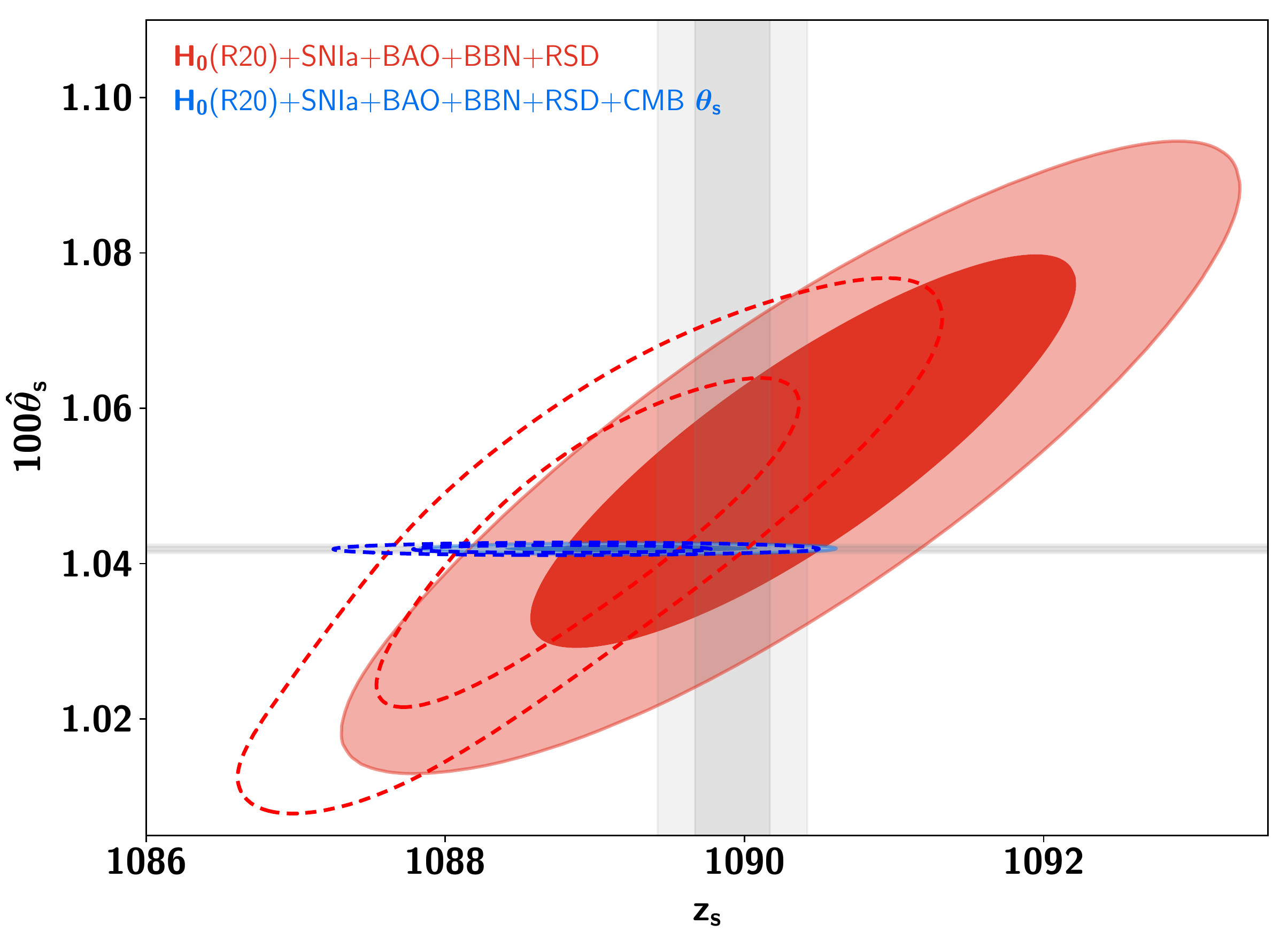}
    \caption{Constraints on $100\hat{\theta}_s\text{--}z_{s}$: Solid contours are for $\fTCDM$ and dashed contours are for $\LCDM$ model; grey bands represent 1$\sigma$ and 2$\sigma$ constraints from CMB Planck 2018 on the acoustic scale and the redshift at recombination \citep{Aghanim:2018eyx}. {In absence of any constraint from the CMB $\theta_{\textrm s}$, i.e the red contours, the $\fTCDM$ theory predicts an acoustic scale a little bit different from $\LCDM$ predictions. This deviation can be understood in terms of slight different evolutions at late time, as clear in Fig.~\ref{Fig:datasets}. However, by including an the precise CMB $\theta_{\textrm s}$ constraint, i.e the blue contours, it similarly fixes the early time evolution of both models in agreement with Planck measurements within $\sim$ 1$\sigma$.}
        }
    \label{Fig:theta_s}
\end{figure}

In order to further examine how $\LCDM$ and $\fTCDM$ fit the Planck CMB measurements, we plot the $\hat{\theta}_{s}$--$z_{s}$ 2D joint contours for both models. As can be seen from Fig.~\ref{Fig:theta_s}, the models provide slightly different values of $\hat{\theta}_s$ at recombination in absence of the CMB $\theta_{\textrm s}$, with a slight preference of $f(T)$ gravity with Planck constraints. On the other hand, both fit well with Planck constraints at the recombination epoch, introduced by including the Planck constraint on the CMB $\theta_{\textrm s}$.In this case  both models have similar early history and deviations in the derived parameters are due to late time evolution.

Despite the similar contours, the small changes still lead to discernible differences in parameters. Using the full likelihood
For $\LCDM$ we obtain $H_0=68.30 \pm 0.77$ km/s/Mpc, which is 3.2$\sigma$ lower than the R20 local measurement $H_0=73.2 \pm 1.3$ km/s/Mpc \citep{Riess:2020fzl}. On the other hand, for $\fTCDM$ we obtain $H_0=70.52 \pm 0.71$ km/s/Mpc using the same dataset, which is 1.8$\sigma$ lower than the local measurement; i.e. the $\fTCDM$ is closer to the R20 measurement than $\LCDM$ by 1.4$\sigma$, while keeping similar $\chi^2_{min}$ for the full dataset. Thus, the analysis shows the Planck constraint on the CMB $\theta_{\textrm s}$ fixes the early history of the two models similarly but allows the exponential IR $f(T)$ theory to better fit with local $H_0$ measurements.

In order to understand how these models take different tracks in the late universe, it is convenient to compare their Friedmann equations, in the matter domination era. For general dynamical dark energy or modified gravity models, we write Friedmann equation
$$E(z) = H(z)/H_0 = \sqrt{\Omega_m (1+z)^3+\Omega_{de} y(z)},$$
where
$$\Omega_{de}=1-\Omega_m \textmd{ and }y(z)=e^{3\textstyle{\int_0^z} \frac{1+w_{de}(z')}{1+z'}dz'}.$$
In the particular case of $\LCDM$, $w_{de}=-1$, we obtain $y(z)=1$. However, in the phantom dark energy with fixed equation of state $w_{de}<-1$, we obtain $y(z)=(1+z)^{3(1+w_{de})}$ which finds $y(0)=1$ and $y(z>0)<1$ in a systematic way as $z$ goes higher. This clearly shows that how phantom energy lowers the expansion rate $E(z)$ at $z>0$ relative to $\LCDM$, which in return increases the angular diameter distance to the last scattering surface $D_A(z_s)=(1+z_s)^{-2} D_L(z_s)$ while keeping the early universe unaltered (in particular the sound horizon $r_s$). Nevertheless the CMB angular scale of the first peak $\theta_s(z_s)=r_s(z_s)/D_A(z_s)$ can be restored to its measured value by accommodating larger $H_0$ value. Similar argument can be applied in the case the exponential IR $\fTCDM$ which effectively imposes dynamical phantom dark energy $-1.2\lesssim w_{de}\lesssim -1$ at late $z\lesssim 8$, see Fig. \ref{fig:Tor_EoS}.
\begin{figure}
    \centering
    \includegraphics[scale=0.4]{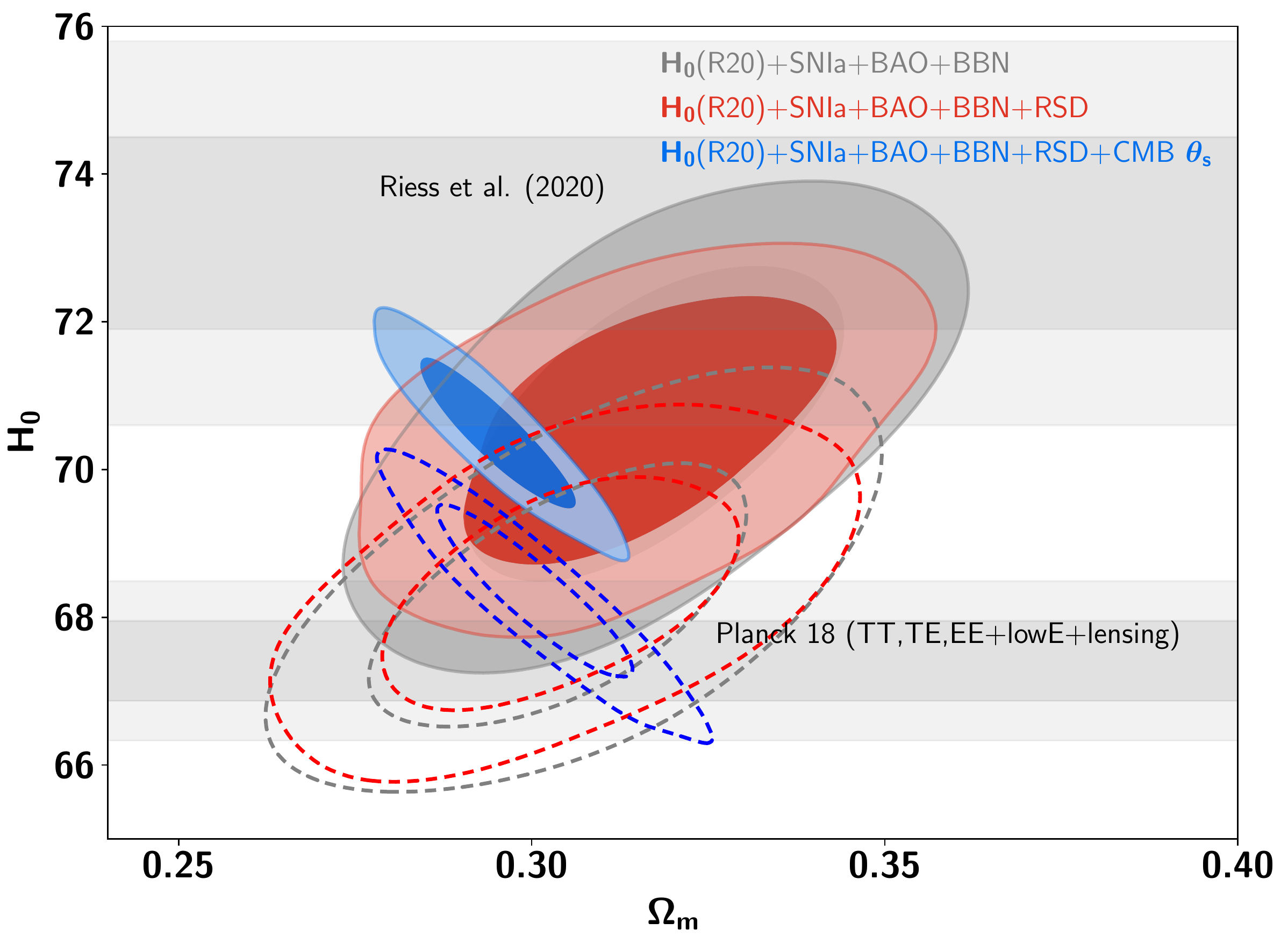}
    \caption{Constraints in the $H_0$--$\Omega_m$ plane: Solid contours are for the $\fTCDM$ model, while dashed contours are for $\LCDM$. The upper grey bands represent 1$\sigma$ and 2$\sigma$ levels of the local measurement of $H_0$ from R20, while the lower bands represent 1$\sigma$ and 2$\sigma$ of Planck 2018 $H_0$ constraints. As is apparent, when adding the CMB $\theta_{\textrm s}$ constraint, $\fTCDM$ tends closer to R20 $H_0$ value (which partially alleviates the $H_0$ tension), while $\LCDM$ model, as expected,  circumscribes
    $H_0$ values similar to  Planck 2018.}
    \label{fig:Omegam-H0}
\end{figure}

Fig.~\ref{fig:Omegam-H0}  shows the
the relevant contours in the $H_0$--$\Omega_m$ plane,
reflecting consistently larger allowable $H_0$
values for the $f(T)$ model relative to $\LCDM$. On the other hand, by adding CMB $\theta_s$ prior, the $\fTCDM$ keeps higher $H_0$ values while slightly decreases not only the mean value of $\Omega_m$ but also decreases its 1$\sigma$ and 2$\sigma$ compared to $\LCDM$ predictions of $\Omega_m$. This shows that the $\fTCDM$ prediction of $\Omega_m$ is more constrained when the CMB $\theta_s$ is considered. These changes still need to be examined with the well measured value $\Omega_m h^2$ known as the geometric degeneracy as provided by the CMB power spectrum.
\subsection{The product $\Omega_m h^2$ and predicted height of CMB peak}\label{Sec:5.4}
We note that the CMB angular acoustic scale $\theta_s$ used in these calculations is  related to the  location of the first acoustic peak of the temperature anisotropy of the CMB power spectrum \citep{Bond:1997wr,Wang:2006ts,Elgaroy:2007bv}. The height of the acoustic peak, on the other hand,
 is constrained by the matter density $\Omega_m h^2$. This was, in principle, left as a free parameter
 in our analysis. But the best fits to both models considered here
 result in values remarkably close to the measured one.
 According to Planck 18 base-$\LCDM$, this physical matter density $\Omega_m h^2=0.1430 \pm 0.0011$ \citep{Aghanim:2018eyx}.
 The full likelihood Base+RSD+CMB $\theta_s$ predicts $\Omega_m h^2=0.1403\pm 0.0053$ for $\LCDM$ --- lower than Planck by $\sim 0.5\sigma$ --- and $\Omega_m h^2=0.1466 \pm 0.0048$ for $\fTCDM$, which is higher than Planck by $\sim 0.7\sigma$.

Thus the two models are interestingly in agreement with Planck within 1$\sigma$ error using mainly late universe observations and
 just the location of the first peak ---
 without involvement of the full CMB power spectrum.
 This is a reflection of a consistency,  once labelled concordance,
 in values of the parameters inferred from different routes.
 The term concordance has come into disuse, in large part in light of the  progressively exacerbating $H_0$ tension. This tension however is alleviated to some extent in the
 context of our present $f(T)$ model without introducing extra parameters.

\subsection{Amplitude of the growth of structure}\label{sec:growthrate}

Another late universe  dataset that is in tension with $\LCDM$-Planck is the cosmic shear measurement of the matter fluctuation by Kilo Degree Survey 450 (KiDs-450). For the flat $\LCDM$ model, the matter amplitude $S_8=\sigma_8 \sqrt{\Omega_m / 0.3}=0.834\pm 0.016$ at $68\%$ CL, as measured by the CMB alone (TT,TE,EE+lowE) \citep{Aghanim:2018eyx}. In contrast, the corresponding value, as measured by KV-450 is $S_8 = 0.737^{+0.040}_{-0.036}$, when using $\LCDM$ with a prior on $H_0$ from direct measurements~\citep{Hildebrandt:2018yau}. The tension between those measurements is thus above 2$\sigma$. Any suggested model to reconcile the early and the late $H_0$ measurements should not strengthen the tension in other measurements like $S_8$.

\begin{figure}
    \centering
    \includegraphics[scale=0.4]{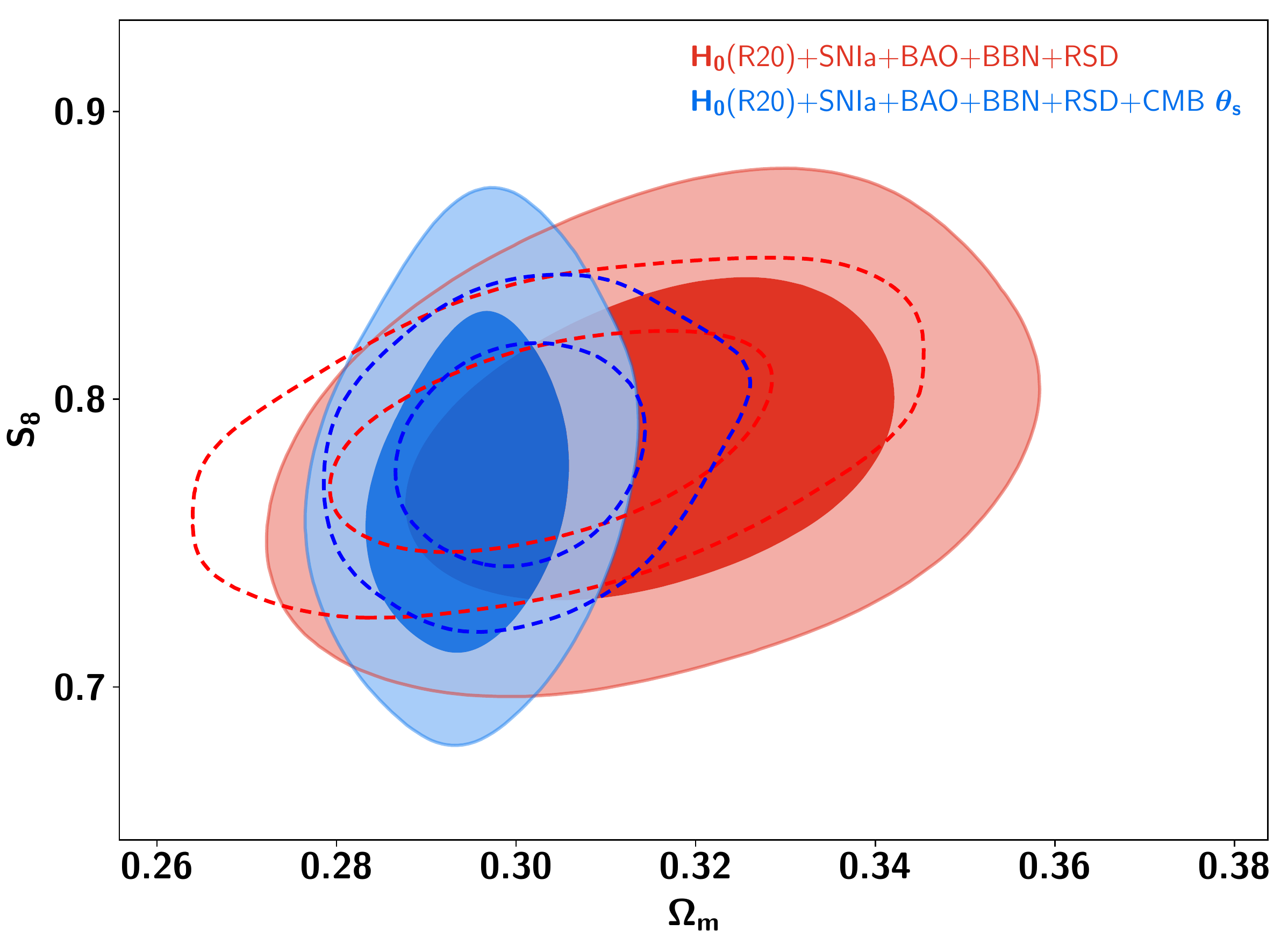}
\caption{Constraints at $68\%$ and $95\%$ CL on $S_8$--$\Omega_m$ plane: Solid contours are for $\fTCDM$ model, while dashed contours are for $\LCDM$. The 2D contour plots show that both models give comparable results with slightly smaller $S_8$ value within the $\fTCDM$ scenario. This shows that the exponential IR $f(T)$ gravity does not worsen $S_8$ tension in comparison to the $\LCDM$ model. However, full investigation requires us to derive the $S_8$ value, at the linear perturbation level of $\fTCDM$ theory, as inferred by CMB powerspectrum too and to compare it with late universe measurements. This is shown in paper II \citep{Hashim:2019b}.}
    \label{fig:Omegam-sigma8}
\end{figure}

Recalling the discussion about the $f(T)$ modification of growth function and the RSD measurements in Subsections \ref{Sec:2.2} and \ref{Sec:4.1.4}, respectively, whereas the growth rate in $f(T)$ scenario is given by Eqs \eqref{eq:growthrate}, \eqref{eq:growthfactor} and \eqref{eq:growth_fn} while the RSD observations measure the product ${f}\sigma_8$. Therefore, by extracting the amplitude $\sigma_8$ and consequently the $S_8$ parameter, we plot the two dimensional $S_8$--$\Omega_m$ plane for both $\LCDM$ and exponential IR $\fTCDM$ models using the joint likelihood without/with CMB $\theta_{\textrm s}$ constraint as obtained in Fig.~\ref{fig:Omegam-sigma8}. It is obvious that the 2D contour plots for both $\LCDM$ and $\fTCDM$ models are compatible within 1$\sigma$ regions.

In more detail, using the Base+RSD data without(with) CMB $\theta_s$ constraint, we respectively obtain $S_8=0.786\pm 0.025$ ($0.782\pm 0.025$) for $\LCDM$ and slightly smaller $S_8=0.785\pm 0.035$ ($0.774^{+0.035}_{-0.041}$) for $\fTCDM$ which are in agreement with the cosmic shear $\LCDM$-base measurement $S_8=0.737^{+0.040}_{-0.036}$ at 68\% CL as provided by KV-450. In addition, both predict almost same $\sigma_8$ values as obtained in Table \ref{Table:datasets}, which indicates that the exponential IR $f(T)$ gravity does not- in principal- worsen the $S_8$-tension.

Although we restrict ourselves to linear matter perturbation in the Newtonian limit (well below horizon scales), a full analysis
is not expected to modify the aforementioned conclusions, as significant modifications in its context would arise on scales
much larger than those affecting than those associated with the $S_8$ tension. In order to investigate the $S_8$-tension within the IR $f(T)$ gravity we need to derive its value as inferred by Planck full CMB and KiDs-450 (or KV-450) cosmic shear, simultaneously, and then we can properly check their consistency. This requires the extension of our analysis to the full linear perturbation effects. We leave such a fuller examination to paper II \citep{Hashim:2019b}.

In sum, at the background level, we find that the $\fTCDM$ theory \eqref{exp-IR} can fit well with different types of observations. It is statistically similar to $\LCDM$ and it can serve as a viable theory of gravity, while providing a framework for reducing tensions between early and late universe with $H_0$ and tentatively does not worsen $S_8$. Although, the exponential IR $f(T)$ theory and $\LCDM$ are conceptually different, they share the same number of free parameters and are statistically similar in viability in terms of the tests undertaken here.

\section{Conclusion}\label{Sec:6}
Late accelerated expansion is a crucial issue of unclear origin in contemporary cosmology. The range of possibilities is unconstrained in such a way as to allow for a cosmological constant, dynamical dark energy or modified gravity as sources for the phenomenon. If one assumes the latter option, IR corrections to gravitational theories may represent a viable scenario; as they modify gravity on cosmic distances while keeping GR predictions fulfilled on smaller
scales, such as the solar system scales where it is very well tested.
In this context, the exponential IR $f(T)=Te^{\beta T_0/T}$ gravity was proposed in Ref.~\citep{Awad:2017yod}, its dynamical phase portrait was examined and shown to account for late time acceleration. Here, we examined in detail the empirical viability of the consequence of
that model on the background dynamics of  cosmic expansion; confronting it with various datasets, covering widely spaced epochs and scales of the Universe.

As the theory does not introduce any extra free parameters compared to  $\LCDM$, it allows for statistical comparison on equal footing. This is unlike  other viable $f(T)$, or modified gravity theories in general, which usually include at least one extra free parameter. This being the
case, one need not resort to such techniques as performing Bayesian information criteria (BIC = $-2 \ln L^{max} + n \ln m$) to obtain well grounded comparisons --- since $n$ (number of parameters) and $m$ (number of data points) are the same for both models, and only the maximum likelihood $L^{max}$ (i.e. $\chi^2_{min}$) affects the results. As $\LCDM$ is already very successful in fitting available data,  modified gravity models that improve on those fits  by adding  tiny modifications through extra parameters are likely to be associated with worse BIC, especially when the number of data points is not so large, c.f. \citep{Xu:2018npu}. Furthermore, marginalizing over extra parameters has the effect of enhancing the $H_0$ tension by broadening the uncertainties of its inferred value rather an actual shift of the $H_0$ mean value \citep{Vagnozzi:2019ezj}. Such issues are avoided here.

We used SNIa, $H_0$, BAO, RSD, BBN, and CMB $\theta_{s}$ to examine the viability of the theory. We evaluated the joint likelihood analysis to find the best-fit values of the four model's parameters $\left\{H_0,~\Omega_b,~\Omega_{c},~\sigma_8 \right\}$ for both $\LCDM$ and $\fTCDM$ models. After
appropriate modification of \verb"CLASS" code with \verb"Monte Python", we ran MCMC samples. Then, we used \verb"GetDist" python package to analyze MCMC chains and get the 2D contour plots at 1$\sigma$ and 2$\sigma$ regions of the different model parameters.

The comparison clearly illustrates that $\fTCDM$ and $\LCDM$ show similar statistical success when confronted with the various datasets. Moreover, while including the accurate CMB constraint $\theta_{\textrm s}$, the $\fTCDM$ theory makes it possible to decrease the $H_0$ tension by 1.4$\sigma$ relative to the $\LCDM$ prediction using the dataset presented in this study, while still giving an age for the universe ($\sim$ 13.76 Gyr), compatible with other astrophysical observations.

The product $\Omega_m h^2$, on which depends the height of the first CMB peak is left free in our analysis, which uses late universe data, in addition to the angular location from $\theta_{\textrm s}$.
Nevertheless, the measured value is obtained within 1-$\sigma$ for both models. This is reflection of a consistency between parameter values obtained from different routes. Once termed concordance, the term has come into disuse, partly due to the progressively serious $H_0$ tension. As mentioned, this tension is less serious in the context of the teleparallel-based cosmology presented here.

The exponential IR $f(T)$ gravity considered here drives the effective equation of state to slip significantly into the phantom regime at lower redshifts, as in the models studied in~\citep{El-Zant:2018bsc}.
Significant deviations from $\LCDM$ are milder and occur later than those associated with the inverse power-law $f(T)$ discussed there however. This allows the exponential IR $f(T)$ to be in less severe tension with BAO measurements, in particular the angular distance, while partially alleviating the $H_0$ tension. In the present study, this alleviation arose as a compromise statistical optimum fitting the various datasets. Larger tensions with the BAO distances may be expected if full resolution between CMB and local $H_0$ measurements is required.

Although we were mainly concerned with constraints arising from the background evolution, we have included
the effect on the growth on linear perturbations on (Newtonian) scales significantly smaller than the horizon, as the effect
turns out to be scale free in the linear regime. In this context, constraints from redshift space distortion suggest
that that the exponential IR $f(T)$ gravity leads to to slightly smaller $S_8 = \sigma_8\sqrt{\Omega_m/0.3}$,
while keeping $\sigma_8$ almost the same as for  $\LCDM$. This indicates that the exponential IR $f(T)$ gravity does not worsen the tension
associated with the normalization of the amplitude of fluctuations.
In a  companion paper we perform a full perturbation analysis and compare the results with the
CMB spectrum.

The IR correction approach is not limited to $f(T)$ teleparallel gravity, which should be primarily seen as an example of how
modified gravity models can may successfully explain the late accelerated expansion by weakening the gravity on the cosmic distances. Although $f(T)$ cosmology is generally simpler to handle mathematically, a major challenge concerns extending its predictions to the non-linear regime of structure formation, ultimately attempting to adopt it to $N$-body simulations, in order to fully test its viability and consequences.

\subsection*{Acknowledgements}
This project was supported financially by the Science and Technology Development Fund (STDF), Egypt. Grant No. 25859. The likelihood analysis presented in this work were done on the Sciama High Performance Compute (HPC) cluster which is supported by the ICG, SEPNet and the University of Portsmouth.
\bibliographystyle{JHEP}

\providecommand{\href}[2]{#2}\begingroup\raggedright\endgroup

\end{document}